\documentclass[aps,pra,showpacs,twocolumn,groupedaddress]{revtex4}
\usepackage{mathtools}
\usepackage{amssymb}
\usepackage{graphics}
\usepackage[dvips]{graphicx}
\usepackage{graphicx}
\usepackage{color}

\newcommand \beq{\begin{eqnarray}}
\newcommand \eeq{\end{eqnarray}}
\def\simge{\mathrel{%
       \rlap{\raise 0.511ex \hbox{$>$}}{\lower 0.511ex \hbox{$\sim$}}}}
\def\simle{\mathrel{
       \rlap{\raise 0.511ex \hbox{$<$}}{\lower 0.511ex \hbox{$\sim$}}}}

\newcommand \la{\raisebox{-.5ex}{$\stackrel{<}{\sim}$}}
\newcommand \ga{\raisebox{-.5ex}{$\stackrel{>}{\sim}$}}

\begin{document}
\title{Transport in ultradilute solutions of $^3$He in superfluid $^4$He}
\author{Gordon Baym,$^{a,b}$ D.\ H.\ Beck,$^a$ and C.\ J.\  Pethick$^{a,b,c}$}
\affiliation{\mbox{$^a$Department of Physics, University of Illinois, 1110
  W. Green Street, Urbana, IL 61801} \\
\mbox{$^b$The Niels Bohr International Academy, The Niels Bohr Institute, University of Copenhagen,}\\
\mbox{Blegdamsvej 17, DK-2100 Copenhagen \O, Denmark}\\	
\mbox{$^c$NORDITA, KTH Royal Institute of Technology and Stockholm University,}\\
\mbox {Roslagstullsbacken 23, SE-10691 Stockholm, Sweden} 
}

\date{\today}

\date{\today}

\begin{abstract}

We calculate the effect of a heat current on transporting $^3$He dissolved in superfluid $^4$He at ultralow concentration,  as will be utilized in a proposed experimental search for the electric dipole moment of the neutron (nEDM).  In this experiment, a phonon wind will generated  to drive (partly depolarized) $^3$He down a long pipe.
In the regime of $^3$He concentrations $\la 10^{-9}$ and temperatures $\sim 0.5$ K, the phonons comprising the heat current are kept in a flowing local equilibrium by small angle phonon-phonon scattering, while they transfer momentum to the walls via the $^4$He first viscosity.  On the other hand, the phonon wind drives the $^3$He out of local equilibrium via phonon-$^3$He scattering.      For temperatures below $0.5$ K, both the phonon and $^3$He mean free paths can reach the centimeter scale, and we calculate the effects on the transport coefficients.  We derive the relevant
transport coefficients, the phonon thermal conductivity and the $^3$He diffusion constants  from the Boltzmann equation.   We calculate the effect of scattering from the walls of the pipe and show that it may be characterized by the average distance from points inside the pipe to the walls. 
The temporal evolution of the spatial distribution of the $^3$He atoms is determined by the  time dependent $^3$He diffusion equation, which describes the competition between advection by the phonon wind and $^3$He diffusion.  As a consequence of the thermal diffusivity being small compared with the $^3$He diffusivity,  the scale height of the final $^3$He distribution is much smaller than that of the temperature gradient.  We
present exact solutions of the time dependent temperature and $^3$He distributions in terms of a complete set of normal modes.

\end{abstract}


\maketitle

\section{Introduction}

The physics underlying the transport properties of mixtures of $^3$He and superfluid $^4$He changes markedly as the concentration of $^3$He varies.  We determine here the transport properties of these mixtures at very low concentrations, $x_3 = n_3/(n_3+n_4)\la10^{-9}$, where $n_3$ and $n_4$ are the $^3$He and $^4$He densities, and low temperatures, $T \,\la\, 0.6$~K, where phonons are the dominant superfluid excitation.  In this case, the phonons are in local thermal equilibrium; their interactions with the $^3$He distort the $^3$He distribution and dominate the $^3$He diffusion.  For concentrations $x_3\, \ga\, 10^{-4}$, the reverse situation holds: the $^3$He are in local equilibrium due to rapid $^3$He--$^3$He scattering and the phonon distribution is distorted due to phonon--$^3$He interactions~\cite{hix}. In the intermediate concentration regime, the phonon and $^3$He distributions are both distorted and must be determined by solving the coupled evolution, or Boltzmann, equations~\cite{BBPII}.  At the highest concentrations, $x_3 \sim 1$\%, Fermi-Dirac statistics for the $^3$He become important~\cite{BBP}.  These transport properties are of interest as an example of a two-component fluid with excitations of comparable energy but very different momenta, and where the  excitations of the two species obey different statistics.

The transport properties of $^3$He in superfluid $^4$He at low concentrations are also important for the proposed experiment~\cite{snsExpt} to measure the neutron electric dipole moment (nEDM) at the Oak Ridge National Laboratory Spallation Neutron Source.  There, the neutron precession frequency will be determined using the absorption of polarized ultracold  neutrons on polarized $^3$He atoms in solution in superfluid $^4$He via the reaction
\beq
{\rm n} + {\rm ^3He} \rightarrow {\rm p} + {\rm t} + 764 \hbox{ keV},
\label{eq:capture}
\eeq
which has a strong spin dependence, since capture proceeds primarily through the spin-singlet channel.  Two key considerations accrue from this choice of detection technique.  In order to maximize the precision with which the precession frequency can be measured, the optimal $^3$He concentration, $x_3\sim 10^{-10}$, corresponds to a capture rate comparable to the decay rate of the neutrons. However, primarily due to wall collisions, the $^3$He will gradually become depolarized.  In order to reduce the background from neutron capture on unpolarized $^3$He, it is crucial to be able to periodically sweep out the $^3$He by means of a heat current \cite{hayden}. In this paper, we calculate both the heat and $^3$He particle currents based on well-established microscopic theory of phonon--phonon \cite{maris} and phonon--$^3$He scatterings \cite{BE67}, as well as the evolution of both the temperature and $^3$He concentration.  

At the concentrations and temperatures of interest in the experiment, in addition to phonon-phonon, phonon-$^3$He and $^3$He--$^3$He scattering, the scattering of both phonons and $^3$He from the walls of the containers can also be important.  Here we extend the solution of the Boltzmann equations in Ref.~\cite{BBPII} to include these effects, in addition to providing some examples for $x_3
\la 10^{-9}$.  For illustration, we consider the effect of a heat current in an essentially one dimensional geometry, with the $^3$He superfluid $^4$He mixture in a long pipe with a diameter of a few cm.  The phonon-wall interactions affect the thermal conductivity as well as the phonon velocity distribution within the pipe; the $^3$He--wall interactions affect the transport of the $^3$He in the presence of a heat current.

This paper is arranged as follows:  Section II describes the basic scattering mechanisms the calculation of transport coefficients from the Boltzmann equation is given in Sec.\ III.  Subsequently, we calculate the temporal and spatial evolution of the temperature (Sec.\ IV) and the $^3$He density (Sec.\ V).  We summarize results in Sec.\ VI.  In Appendix A we analyze the transport when scattering of phonons is predominantly from the walls of the pipe, and in Appendix B we solve analytically the equation for the temporal evolution of the $^3$He concentration.

\section{Phonon and $^3$He relaxation}

We begin by considering the  relevant microscopic relaxation mechanisms (detailed in Ref.~\cite{BBPII}), first for the phonons.  The 
momentum-dependent mean free path of a phonon of momentum $q$ scattering against the $^3$He,
\beq
  \ell_{ph3}(q) = \frac{s}{\gamma_q} = \frac{4\pi n_4}{x_3 J}\frac{1}{q^4},
  \label{k1}
\eeq  
is typically greater than 1 km for $x_3\, \la\, 10^{-9}$ and $T \sim 0.5$~K ~\cite{BBPII}; here $\gamma_q$ is the corresponding scattering rate, $s$ is the phonon velocity, and $J$ is an angle-integrated rate constant.  Therefore, phonon--wall and phonon--phonon scatterings determine the phonon distribution.  As discussed in Refs.~\cite{BBPII} and~\cite{BM}, rapid, small angle phonon-phonon scattering establishes thermal equilibrium along phonon `rays,' i.e., given directions in momentum space, 
with the distribution
\beq
  n_{\vec q}^{le}  = \frac{1}{e^{( sq- \vec q \cdot \vec v_{ph})/T(\vec r)} -1},
\eeq
where $T(\vec r)$ is the local temperature and $v_{ph}$ is the mean phonon drift velocity.     Large angle phonon-phonon scattering, 
either in a single event or a succession of small-angle processes, is slower and gives rise to the phonon first viscosity,  
\beq
       \eta_{ph} = \frac15 \frac{TS_{ph}}{s}\ell_{visc}.
     \label{eta}
\eeq
where $S_{ph}$ is the phonon entropy density and $\ell_{visc}$ is the viscous mean free path.  At a pressure of $0.1$~bar,
\beq
\ell_{visc}& \simeq &\frac{3.2\times10^{-3}}{T_K^5} {\rm cm},
\label{eqn:ellvisc}
\eeq
to a good approximation~\cite{maris,greywall},
where $T_K$ is the temperature in Kelvin; at $T$= 0.45 K, $\ell_{visc} \simeq $ 0.17 cm.

 In the presence of a heat flux, $\vec Q=T S_{ph}\vec  v_{ph}$,  small-angle phonon-phonon scattering keeps the phonons in local thermal equilibrium, where the mean phonon drift velocity, $v_{ph}$, is 
 \beq
T S_{ph}\vec  v_{ph} = -K_{ph}\nabla T.
\label{eqn:Q}
\eeq
The thermal conductivity of the phonons, $K_{ph}$, can, at low concentrations, be written as~\cite{BBPII,greywall}, 
\beq
K_{ph} &=& \frac{5}{8}\frac{s S_{ph} R^2}{\ell_{eff}},
\label{eqn:Kph}
\eeq
in a pipe of radius $R$ and where $\ell_{eff}$ is the effective mean free path
\beq
\frac{1}{\ell_{eff}} &=& \frac{1}{\ell_{visc}} + \frac{16}{5R}.
\label{eqn:leff}
\eeq
The second term represents scattering of the phonons on the walls; the numerical coefficient 16/5 is chosen to give the correct
Casimir limit.  In this limit, $\ell_{visc}$ large compared to the pipe diameter (see Appendix A), the phonon thermal conductivity assumes the Casimir form~\cite{casimir} 
\beq
K_{ph,Casimir} &=& 2Rs S_{ph}.
\label{eq:KCasimir}
\eeq
In the opposite limit, $\ell_{visc} \ll R$, the thermal conductivity becomes
\beq
K_{ph,visc} &=& \frac{5}{8}\frac{s S_{ph} R^2}{\ell_{visc}}.
\label{eq:Kvisc}
\eeq
The $^3$He contribution to the overall heat flux is negligible at low $x_3$~\cite{hix,BBPII}.

The mean free path of a $^3$He scattering on unpolarized $^3$He is~\cite{BBPII} 
\beq
\ell_{33}=\frac{1}{(n_3/2) \sigma_{\rm 33}}  =\frac{8.66\times 10^{-8}}{x_3} {\rm cm},
\label{l33} 
\eeq
where $\sigma_{33}$ is the corresponding cross section.  Thus for $x_3 < 10^{-9}$, one has $\ell_{33}\, \ga \,1$~m;
 we therefore neglect $^3$He--$^3$He scattering.
On the other hand, the mean free path for $^3$He scattering on phonons~\cite{BBPII},
\beq
  \ell_{3ph} &=&   \frac{\sqrt3}{2J}  \left(\frac{n_4}{S_{ph}}\right)^2   \frac{m^{*1/2} s^2}{T^{3/2}} 
    \nonumber \\
      &=& 0.077 \left(\frac{0.45\,K}{T}\right)^{15/2} \, {\rm  cm},
 \label{l3ph}   
 \eeq
where $m^*$ is the $^3$He effective mass in superfluid $^4$He, is small compared to the pipe diameter for $T\, \ga \,0.3$~K.  Thus the dominant process for bringing the $^3$He toward equilibrium for temperatures of interest in the experiment is scattering against phonons.  In the next section, we outline the calculation of the $^3$He transport coefficients; more details can be found in Ref.~\cite{BBPII}.

\section{$^3\mbox{\boldmath He}$ Boltzmann equation and transport coefficients}

The $^3$He Boltzmann equation has the general form
 \beq
&&\frac{\partial f_{\vec p}}{\partial t}  +\frac{\vec p}{m^*}\cdot \nabla_r f_{\vec p} \nonumber\\
&& =\sum_{p',q,q'}{\cal T} \left[f_{\vec p\,'}n_{\vec q\,'}^{le}(\vec r\,)(1+n_{\vec q}^{le}(\vec r\,)) \right. \nonumber \\
&&\hspace{20mm}\left. -f_{\vec p}\,n_{\vec q}^{le}(\vec r\,)(1+n_{\vec q\,'}^{le}(\vec r\,)\right] 
\nonumber\\ &&- \frac{\delta f_{\vec p} - \beta f_p^0 v_3}{\tau_{33}}  - \frac{\delta f_{\vec p} - \beta f_p^0 v_{3}}{\tau_{3ws}}- \frac{\delta f_{\vec p}}{\tau_{3wd}}.
\label{eq:3HeB}
\eeq
Here $f_{\vec p}$ is the $^3$He distribution function, 
\beq
f_p^0 = e^{-\beta(p^2/2m^* - \mu_3)}
\eeq
is the equilibrium distribution function, and we write the deviations from local equilibrium as
\beq
\delta f_{\vec p} = f_{\vec p} - f_p^{le0},
\eeq
where 
\beq
f_{\vec p}^{le0}  = e^{-(p^2/2m^*  -\vec p\cdot \vec v_{ph}- \mu_3(\vec r))/T(\vec r)}
\eeq
is the local equilibrium distribution function, i.e., the distribution towards which collisions with phonons drive the $^3$He.
The first term on the right represents the scattering on the phonons, the second term the scattering from other $^3$He (numerically insignificant for the concentrations of interest) and the last two terms isotropic diffuse (d) and specular (s) scattering of the $^3$He from the walls. In the term describing collisions with phonons, $\vec p$ and $\vec p\,'$ are the initial and final $^3$He momenta, respectively, and
 $\vec q$ and $\vec q\,'$ are the corresponding phonon momenta.
The phonon--$^3$He scattering kernel is ${\cal T} \equiv |\langle p'q'|T|pq\rangle|^2 2\pi\delta( p^2/2m^*+sq-p'^2/2m^* - sq')$, and momentum conservation,
$\vec p\,' +\vec q\,'= \vec p + \vec q$, is understood in the collision term.  

   In order to calculate the effects of a phonon wind on the $^3$He we solve Eq.~(\ref{eq:3HeB})
for $\delta f_{\vec p}$.  
On the left side of the Boltzmann equation, we approximate the distribution by its local equilibrium form, $f_p^{le0}$. 
We neglect the contribution from $\partial v_{ph}/\partial z$ because of the relatively small temperature gradient (see Eq.~(\ref{eq:DTterm}) below), while the gradient of  $\beta$ in this term gives a second order contribution, which we neglect.  The left side of the Boltzmann equation is then
\beq 
 \frac{\partial f_p^{le0}}{\partial z} = 
    \left[\frac{p^2}{2m^*} - \frac32 T\right] f_p^0 \frac{1}{T^2}\frac{\partial T}{\partial z}
     + f_p^0 \frac{1}{n_3} \frac{\partial n_3}{\partial z} .
\eeq

 On the right side we 
write
\beq
\delta f_{\vec p} \equiv \beta f_p^0 p_z w_p;
\eeq
as shown in Ref.~\cite{BBPII} the $^3$He-phonon collision term is diagonalized by expanding $w_p$ in Sonine polynomials~\cite{BBPII}.

   To solve for $w_p$ we multiply the Boltzmann equation by $p_z$ and integrate over all $\vec p$.  
Noting that the distortion $\delta f_{\vec p}$ is proportional to $p_z$, we see first that on the right side of the Boltzmann equation the two term $\propto \delta f_{\vec p} - \beta f_p^0 v_{3}$ do not contribute, since both the $^3$He--$^3$He scattering and the specular $^3$He scattering from the walls conserve momentum in the z direction.  Following Eq.~(81) of Ref.~\cite{BBPII} for the phonon-$^3$He scattering, we see that the remaining terms on the right side comprise
\beq
-\frac{\beta\Gamma}{3 m^*}\left[ \delta f_{\vec p} - \beta f_p^0 p_z v_{ph} \right ] - \frac{\delta f_{\vec p}}{\tau_{3wd}}.
\eeq
With the inclusion of the recoil effect in the phonon--$^3$He scattering to lowest order (see the Appendix of Ref.~\cite{hix} and Sec. 6 of Ref.~\cite{BBPII}), the solution of the Boltzmann equation is 
\beq
  \delta f_{\vec p} = \tau_3^\prime\frac{p_z}{m^*} \left[\frac{\beta^2\Gamma_{rec}}{3} f_p^0 v_{ph}- \frac{\partial f_{p}^{le0}}{\partial z} \right],
\eeq
where
\beq
\frac{1}{\tau_3^\prime} \equiv \frac{\beta\Gamma_{rec}}{m^*} + \frac{1}{\tau_{3wd}}
\eeq
is the effective $^3$He scattering rate, including both scattering from phonons, encoded in $\Gamma_{rec}$, and diffuse scattering from the walls of the pipe.

  Integrating the Boltzmann equation, Eq.~(\ref{eq:3HeB}), over $\vec p\,$ we recover the continuity equation
\beq
\frac{\partial n_3}{\partial t} + \nabla \cdot \vec j_3 = 0,
\label{eq:continuity}
\eeq
with the $^3$He particle current given by
\beq
\vec j_3 = \nu \int \frac{d^3p}{\left ( 2 \pi \right )^3} \frac{\vec p}{m^*} \delta f_{\vec p},
\label{eq:j3}
\eeq
where $\nu$ is the number of spin degrees of freedom of the $^3$He: 1 for a fully polarized sample and 2 in the unpolarized case.
It is straightforward to evaluate the current (including the effect of recoil in the phonon--$^3$He scattering)
\beq   
  j_3 = n_3 v_{ph} - D_{rec} \frac{\partial n_3}{\partial z}
 - D_{T,rec}  \frac{\partial T}{\partial z}
\label{eqn:j3}
\eeq
where the $^3$He diffusion constant, including recoil corrections, is~\cite{BBPII} 
\beq
D_{rec} = 3 \xi\left(R,\zeta\right ) T^2/\Gamma_{rec},
\label{eqn:D}
\eeq
and the ``thermoelectric'' coefficient is
\beq
D_{T,rec}  = 3\xi\left(R,\zeta\right ) T n_3/\Gamma_{rec}.
\label{eqn:DT}
\eeq
In these expressions the basic forms of $D$ and $D_T$ are modified by the wall scattering factor
\beq
\xi\left(R,\zeta\right )= \left ( 1+\frac{3 m^*}{\beta\Gamma_{rec}}\frac{\zeta}{\tau_{3wd}\left( R\right)} \right )^{-1},
\label{eq:xi}
\eeq
where 
\beq
\zeta = \frac{\tau_{3wd}^{-1}} {\tau_{3wd}^{-1} +\tau_{3ws}^{-1}} 
\eeq
is the fraction of the $^3$He wall scattering rate that is diffuse. 

\begin{figure}[t]
\includegraphics[width=8.5cm]{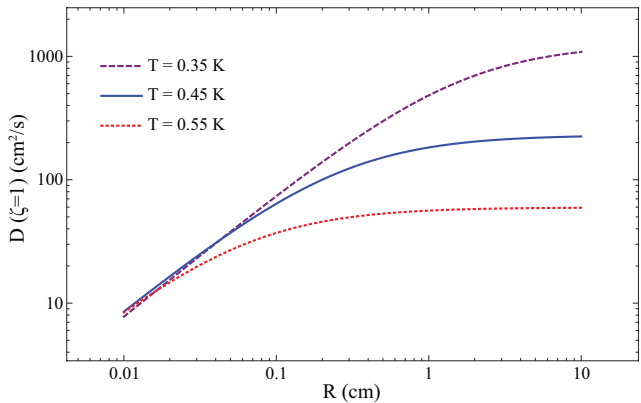}
\caption{(color online) The diffusion constant, $D_{rec}$, Eq.~(\ref{eqn:D}) as a function of the pipe radius, $R$, for $T=0.35$ K (purple, dashed), $T=0.45$ K (blue, solid), and $T=0.55$ K (red, dotted).  We assume here that the $^3$He wall scattering is diffuse, $\zeta=1$.}
\label{fig:DR}
\end{figure}

\begin{figure}[t]
\includegraphics[width=8.5cm]{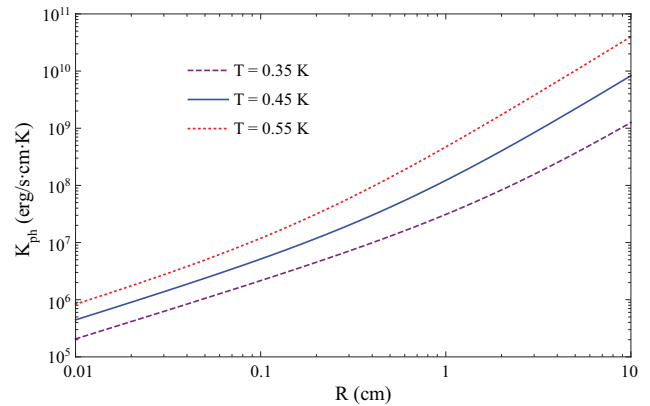}
\caption{(color online) The phonon thermal conductivity, $K_{ph}$, from Eq.~(\ref{eqn:Kph}) for $T = 0.35$~K (dashed), $0.45$~K (solid) and $0.55$~K (dotted) as a function of pipe radius.   For $R\ll \ell_{visa}$ the conductivity rises as $R$ (Eq.~(\ref{eq:KCasimir})), while for $R \gg \ell_{visc}$, it rises 
as $R^2$ (Eq.~(\ref{eq:Kvisc})).}
\label{fig:KPh}
\end{figure}

   To see the effect of scattering of $^3$He from the walls,  we take the total wall scattering rate to be simply 
\beq
\frac{1}{\tau_{3wd}\left( R\right)} = \frac{\bar v_3}{{\cal D}_{eff}},
\label{eq:3HeWallRate}
\eeq
where $\bar v_3 = \sqrt{3T/m*}$ is the mean $^3$He thermal velocity, and
${\cal D}_{eff}=2R/3$ [see Eq.~(\ref{deff})] is the effective average distance from an interior point to the wall of an infinitely long pipe of radius 
$R$ entering the transport \cite{knudsen}.  The effect  of wall scattering on the diffusion constant, $D_{rec}$, for example, is shown in Fig.~\ref{fig:DR} for $\zeta = 1$.   As we see, the effect becomes more important for lower temperatures because of the decrease in the phonon density.  As $R\to\infty$, $D_{rec}$ approaches the result without wall scattering, which  in the vicinity of the operating temperature regime of the nEDM experiment, $T \sim 0.45$~K,  is
\beq
D_{rec} \cong  \frac {0.88}{T_K^7}\ \hbox{cm$^2$/s}.
\eeq
We also show, in Fig.~\ref{fig:KPh}, the corresponding effect of phonon--wall scattering on the phonon thermal conductivity,
Eq.~(\ref{eqn:Kph}). 


At temperatures of approximately 0.3K and below, the mean free path of a
$^3$He quasiparticle in the bulk medium is greater than 1 cm, which is
comparable to the radii of the pipes considered for the nEDM experiment.  In
this regime, the $^3$He quasiparticles can lose momentum not only by
collisions with phonons but also directly to the walls of the pipe.
This situation is analogous to the (Knudsen) flow of a low-density gas in
response to a pressure gradient, except that in the
dilute helium solutions the force driving the flow of $^3$He has two
components, one due to the $^3$He pressure gradient and another due to
the collisions with phonons.   When the $^3$He distribution function
is stationary, the two contributions to the force are
equal and opposite.  The $^3$He distribution function  in principle depends not only
on direction in momentum space but also on the radial coordinate in the
pipe. However, the time scales for 
smoothing out the radial dependence via diffusion or ballistic transport are of order
several milliseconds, and therefore short compared with the overall evolution time
scales of the system.  A detailed study of this regime lies outside the scope of the
present article.

\section{Time evolution of the temperature}

We now estimate, using the heat diffusion equation. the timescale for heating the fluid.   We assume, as above, that the heat is carried by the phonons (see also Eq. (89) of Ref.~\cite{BBPII} and discussion following) and that the relative temperature variation and, hence, the variation of $K_{ph}$ is small, so that
\beq
\frac{\partial \epsilon}{\partial t} = 3 S_{ph} \frac{\partial T}{\partial t} = -\nabla\cdot \vec  Q = K_{ph} \nabla^2 T,
\eeq
where $\epsilon$ is phonon energy density.   Within a few scattering times after the application of  heat at one end of the pipe ($z=0$), the temperature there is approximately fixed at
 $T_0+\Delta T$.  We assume that the temperature at $z=L$, the other end of the pipe, is kept constant at temperature $T_0$ by a refrigerator.
 
To solve the heat diffusion equation it is sufficient to consider the average of the temperature, $T(z,t)$, over the cross-section of the pipe, thus avoiding having to take into account details of the counterflows within the pipe.    The solution is given in terms of the modes in the pipe that vanish at $z=0$ and $L$,  
$\sin k_\nu z$, where $k_\nu = \nu \pi/L$, with $\nu$ here a positive integer:
\beq
  T(z,t) = T_0 + \Delta T(1-z/L)  + \sum_{\nu\ne 0} c_\nu e^{-D_{th}k_\nu^2 t} \sin k_\nu z .\nonumber\\
\eeq
We denote the thermal diffusivity by
\beq
 D_{th} = K_{ph}/3S_{ph},
\eeq 
and recognize $3S_{ph} = (2 \pi^2/15) \left ( T/s \right)^3$ as the $^4$He specific heat.   The condition that $T(z,t=0) = T_0$ except immediately at $z=0$, implies that the mode weights are given by
$c_\nu = -2\Delta T/\nu \pi$.   The characteristic time, $\tau_{th}$, to set up a steady state phonon wind is essentially that of the $\nu=1$ mode, 
 \beq
 \tau_{th} = \frac{1}{D_{th}k_1^2}=\frac{L^2K_{ph}} {3\pi^2S_{ph}}.
 \eeq
For typical conditions in the experiment, 5 mW of heat in a 3~cm diameter, 100~cm long pipe at $T = 0.45$~K,
the phonon thermal conductivity is $2.4\times 10^8$~erg/s$\cdot$cm$\cdot$K, $\tau_{th} \sim 11$~ms, and $\Delta T = 3$~mK.

\section{Time evolution of the $^3\mbox{\boldmath He}$ concentration}

To begin examining the $^3$He concentration, we consider its steady-state distribution in the presence of a heat current or phonon wind.  Because, as we shall see below, the term involving $D_T$ is relatively small for low concentrations, the condition that the $^3$He particle current, Eq.~(\ref{eqn:j3}), vanishes, is simply
\beq
D_{rec}\frac{\partial n_{3}}{\partial z} = v_{ph}n_{3},
\label{eq:sssimple}
\eeq
which has the solution
\beq
    n_3(z)= \tilde n_3 e^{z/h}  \equiv n_{3,\infty}(z),
\label{eq:ss}
\eeq
where we define the scale height, $h = D_{rec}/v_{ph}$ (for the example parameters above, $D_{rec}=225$~cm$^2$/s, $v_{ph}=17$~cm/s and $h=13$~cm), and
\beq
\tilde n_3 = \frac{n_0}{e^{L/h}-1}\frac{L}{h},
\eeq
with $n_0$ the initial uniform $^3$He density. We note that the relative size of the term involving $D_{T,rec}$  is simply the ratio of the scale heights of the concentration and the temperature,
\beq
\frac{D_{T,rec} \left | \partial T/\partial z\right |}{D_{rec}\left | \partial n_3/\partial z\right |} =  \frac{ \left |\partial \ln T/\partial z\right |}{\left | \partial \ln n_3/\partial z\right |}=\frac{D_{rec} S_{ph}}{K_{ph}},
\label{eq:DTterm}
\eeq
about 1/1000 for the example parameters given above.

In the nEDM experiment, the $^3$He in the system depolarizes in time, primarily due to interactions with walls.  The depolarized $^3$He will be removed by a phonon wind before the system is recharged with more highly polarized $^3$He.  As above, we consider the simple situation of a long pipe with a heater at $z=0$ and closed ends.  The evolution of the $^3$He is governed by a competition between two processes: the phonon wind, which were it to act alone would push all the $^3$He to the downstream (large $z$) end of the pipe,  and diffusion of the $^3$He, limited by scattering with the phonons, which allows the $^3$He to drift back towards smaller $z$.   

This evolution of the $^3$He concentration in the presence of a phonon wind is described by the diffusion equation resulting from Eq.~(\ref{eq:continuity}),
\beq
   \frac{\partial n_3(z,t)}{\partial t} +  v_{ph}\frac{\partial n_3}{\partial z} - D_{rec} \frac{\partial^2 n_3}{\partial z^2} =0,
   \label{diffeq}
\eeq
where we have dropped the $D_T$ term in Eq.~(\ref{eqn:j3}).  Once a steady phonon wind, with a small temperature gradient, is established,
we may neglect the temperature dependence of $D_{rec}$ and take $v_{ph}$ and $D_{rec}$ to be constant. 
For a pipe with a large length to diameter ratio, we may
treat the problem as one dimensional, averaging over its cross section as we did above for the heat flow.
The boundary conditions are that the $^3$He current, $j_3$, Eq.~(\ref{eqn:j3}), vanishes at the two ends of the pipe,
\beq  
   \frac{\partial n_3}{\partial z} =  \frac{n_3}{h} \quad\quad (z=0,L).
\label{bc}   
\eeq 

To solve Eq.~(\ref{diffeq}) with constant $v_{ph}$ and $D$, we write the $^3$He density as  $e^{z/2h}\hat n(z,t)$ and decompose $\hat n(z,t)$ as a sum of time dependent modes
$\hat n_\nu(z,t)$  periodic in $2L$ (see Appendix B):
\beq
   n_3(z,t) &=& e^{z/2h}\sum_{\nu=0}^\infty \hat n_\nu \left( z,t \right ),
    \label{expansion0}
\eeq
where $\hat n$ satisfies the boundary condition
\beq  
   \frac{\partial \hat n}{\partial z} =  \frac{\hat n}{2h} \quad\quad (z=0,L).
\eeq
The spatial parts of the mode functions $\hat n_\nu \left( z,t \right)$ are the complete orthonormal set 
\begin{equation}
\phi_\nu(z) =  
\left\{ 
\begin{array}{lr}
\frac{{\displaystyle e^{z/2h}}}{{\displaystyle \left[ h\left( e^{L/h} - 1 \right )\right]^{1/2}}},& \nu = 0\\ \\
\alpha_\nu \left ( \cos k_\nu z + \frac{1}{2hk_\nu}\sin k_\nu z \right ),& \nu \ge 1
\end{array}
\right. 
\end{equation}
with  $k_\nu =\pi \nu/L$ and $\alpha_\nu = \left[ \left(L/2\right) \left( 1+ 1/(2hk_\nu)^2 \right) \right]^{-1/2}$.  The time dependence of the modes is $e^{-t/\tau_\nu}$ where
\beq
   \frac{1}{\tau_\nu} = \left\{ 
\begin{array}{lr} 0,& \nu = 0\\ \\
  k_\nu^2 D+v_{ph}^2/4D  = \left(k_\nu^2 + 1/4h^2 \right)D, & \nu \ge 1.
\end{array}
\right.  
\label{eq:tauNu}
\eeq
The solution of Eq.~(\ref{diffeq}) for an initially uniform density $n_{3,0}$ is then
\beq
   n_3(z,t) = n_{3,0}\,e^{z/2h}\sum_{\nu=0}^\infty c_\nu \phi_\nu \left( z \right ) e^{-t/\tau_\nu},
   \label{expansion}
\eeq
where 
\beq
c_\nu &=& \int_0^L \left(n( z,0 \right)/n_3^0)\phi_\nu\left( z \right)dz\nonumber \\
&=& \left\{
\begin{array}{lr}
\frac{{\displaystyle L}}{{\displaystyle \left[ h\left( e^{L/h} - 1 \right )\right]^{1/2}}},&  \nu = 0\\ \\
\frac{{\displaystyle 8h \alpha_\nu}}{{\displaystyle 1+(2hk_\nu)^2}} \left( 1+ \left( -1 \right )^{\nu + 1} e^{-L/2h} \right).& \nu \ge 1
\end{array}
\right. 
\label{eq:c}
\eeq

As we show in Appendix B, the general solution may be written in compact form in terms of a Green's function
\beq
  \hat n\left( z,t \right) = \int_0^L {\cal G} \left( z,z',t \right ) \hat n \left( z',0 \right ) dz',
\eeq
where
\beq
{\cal G}\left( z,z',t \right )=\sum_{\nu=0}^\infty \phi_\nu \left( z \right)\phi_\nu \left( z' \right) e^{-t/\tau_\nu} \theta \left( t \right).
\eeq

\begin{figure}[t]
\includegraphics[width=8.5cm]{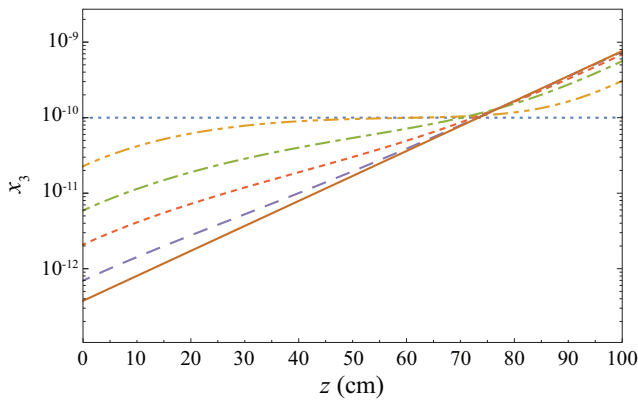}
\caption{(color online) The $^3$He concentration, $x_3$, from the solution of Eq.~(\ref{diffeq}) as a function of $z$, the distance along the pipe, for various times: $t = 0$ (dotted), 1 (dash double dot), 3 (dash dot), 5 (short dash), 8 (long dash) and 20 s (solid).  The result is shown for typical parameters in the nEDM experiment: $x_{3,0}=10^{-10}$ and $5$ mW of heat into a 3 cm diameter, 100 cm long pipe at a nominal temperature of $0.45$~K.}
\label{fig:x3ofz}
\end{figure}
\begin{figure}[t]
\includegraphics[width=8.5cm]{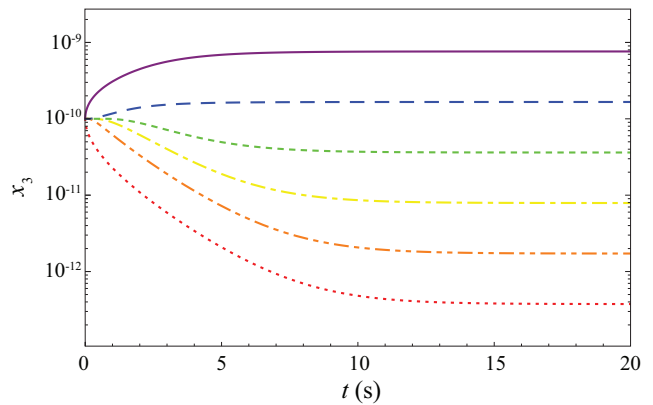}
\caption{(color online) The $^3$He concentration, $x_3$, from the solution of Eq.~(\ref{diffeq}) as a function of time, $t$, for various positions along the pipe: the curves correspond to $z = 0$ (dotted), 20 (dash double dot), 40 (dash dot), 60 (short dash), 80 (long dash) and 100 cm (solid).  Note that it takes about 5.5 time constants, $\tau_1$, for the distribution at the hot end of the pipe ($z=0$) to reach equilibrium.  The result is shown for typical parameters in the nEDM experiment: $x_{3,0}=10^{-10}$, and $5$ mW of heat into a 3 cm diameter, 100 cm long pipe at a nominal temperature of $0.45$~K, for which $\tau_1=1.8$~s.}
\label{fig:x3oft}
\end{figure}

\begin{figure}[t]
\includegraphics[width=8.5cm]{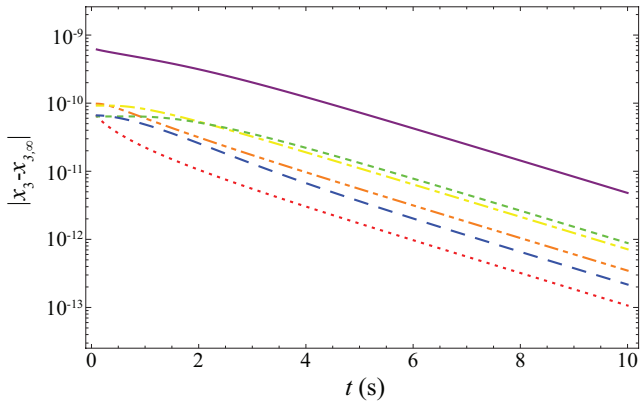}
\caption{(color online) The absolute value of the $^3$He concentration difference, $\left| x_3-x_{3,\infty}\right|$, from the solution of Eq.~(\ref{eq:continuity}) as a function of time, $t$, for various positions along the pipe: the curves correspond to $z = 0$ (dotted), 20 (dash double dot), 40 (dash dot), 60 (short dash), 80 (long dash) and 100  cm (solid).  We note that the lowest mode, corresponding to the time constant $\tau_1$, dominates the time evolution after a few seconds.    The result is shown for typical parameters in the nEDM experiment: $x_{3,0}=10^{-10}$, and $5$ mW of heat into a 3 cm diameter, 100 cm long pipe at a nominal temperature of $0.45$~K, for which $\tau_1=1.8$~s.}
\label{fig:deltax3oft}
\end{figure}

The $z$ and $t$ dependences of $x_3$ are shown in Figs.~\ref{fig:x3ofz} and \ref{fig:x3oft}, respectively, for the case of the uniform initial distribution and typical experimental values (5 mW heat into a 3 cm diameter, 100 cm long pipe at $T=0.45$~K).   As the figures illustrate, the concentration scale height, $h$, is substantially smaller than the pipe length.  Figure~\ref{fig:deltax3oft} plots the difference between $x_3$ and its steady state value for several points along the pipe, showing that, after a few seconds, the lowest mode, with $\tau_1 = 1.8$~s, dominates the time evolution throughout the pipe.

The results for the evolution of the $^3$He concentration presented here are equally applicable to a heat flush experiment being carried out at Harvard at natural $^3$He concentration \cite{flush}.  There one must use the more general phonon thermal conductivity as derived in \cite{BBPII}; phonon-wall scattering in this regime plays a negligible role.

\section{Summary}

We have calculated the transport properties of dilute mixtures, $x_3 \,\la \,10^{-9}$, of $^3$He in superfluid $^4$He at temperatures around 0.5 K where phonons are the dominant excitations of the superfluid.  In this regime, we considered a simple one dimensional geometry (a pipe), a heat current generates a phonon wind with phonons in local equilibrium corresponding to the temperature at that point in the pipe.  On the other hand, phonon scattering distorts the $^3$He distribution from equilibrium.  Starting from the known phonon--phonon and phonon--$^3$He scattering, we calculate the transport coefficients from the Boltzmann equation.  We show that, in the presence of a heat current which generates a temperature scale height much larger than the length of the pipe (i.e., a small relative temperature gradient), the scale height for the $^3$He concentration can be much less than the pipe length.  This leads to a large decrease in concentration at the hot boundary and a corresponding increase at the cold end.  For temperatures below $0.5$~K the mean free paths of both the phonons and the $^3$He can reach the centimeter scale; in these cases, scattering from the walls of the container becomes important.  Finally, we calculate the timescales associated with the evolution of both the temperature and concentration distributions; because of the large superfluid thermal conductivity, the thermal timescales are on the ms scale, whereas the corresponding scale for evolution of the concentration is on the scale of seconds.

\section*{Acknowledgements}

This research was supported in part by NSF Grants PHY-1205671 and PHY-1305891.   GB is grateful to the Aspen Center for Physics, 
supported in part by NSF Grant PHY-1066292, and the Niels Bohr International Academy where parts of this research were carried out, and he thanks Hiroshi Fukuyama for enlightening discussions.  DB thanks Caltech, under the Moore Scholars program, and CJP thanks Andrew Jackson for helpful comments.

\appendix
\section{Phonon drift velocity in the ballistic limit}
\label{sec:AppendixA}

Here we derive the spatial dependence of the phonon drift velocity when the phonon-phonon scattering mean free path for large angle scattering, Eq.~(\ref{eqn:ellvisc}), is much larger than the pipe radius, $R$.   For the pipe geometry considered in the text, this condition holds at a temperature $T \,\la\, 0.3$~K.
We assume that the pipe axis is in the z direction, and that the transverse coordinates are $x$ and $y$.  We also assume that a phonon striking the cylinder wall is diffusively reflected, with a  distribution of final momenta given by the local temperature, $T(z)= T_0+T'z$, where $T' <0$ is the temperature gradient.  Then $n_{\vec q\,}(\vec r\,)$, the number of phonons of momentum $\vec q$ at point $\vec r$, is given by the equilibrium distribution $n^0_q(z')$ at the point $\vec r\,' = (x',y',z')$ on the pipe wall where the phonon at $\vec r$ originated:
 \beq
    n_{\vec q\,}(\vec r\,) = \frac{1}{e^{sq/T(z'(\vec q\,))} -1} \simeq n_q^0  - z'q\frac{T'}{T_0}\frac{\partial n_q^0}{\partial q},
    \label{nz}
\eeq 
to lowest order in $T'$, where $n_q^0$ is the equilibrium distribution.  
    
The point of origin is determined by simple geometry, namely, $\vec r - \vec r\,' = \hat q \,\cal D$, where ${\cal D} = |\vec r - \vec r\,'|$.   We measure $\hat q$ in polar coordinates $\theta_q$ and $\phi_q$.  
Then $z' = z - {\cal D}\cos\theta_q $, $x = x' - {\cal D}\sin\theta_q\cos\phi_q$, and $y' =y-{\cal D}\sin\theta_q\sin\phi_q$.  Using $x'^2 + y'^2 = R^2$ on the cylinder wall, we have then
\beq
\sin^2\theta_q\, {\cal D}^2  - 2\rho {\cal D}\sin\theta_q\cos(\phi_q-\phi_r) - R^2 +\rho^2 = R^2, \nonumber\\
\eeq 
where $\phi_r$ is the azimuthal angle of $\vec r$ and $\rho = \sqrt{x^2+y^2}$.
The solution is 
\beq
{\cal D} = \frac{1}{\sin\theta_q}\left[\rho\cos(\phi_q -\phi_r)+\sqrt{R^2-\rho^2\sin^2(\phi_q-\phi_r)}  \right];\nonumber\\
\label{d}
\eeq
without loss of generality we take $\phi_r=0$.  

   The local phonon flow velocity $v_{ph,z}(\rho)$ is given by the local total momentum flux density in the z direction divided by
the normal mass density of the phonons, $\rho_{ph}$,
\beq
v_{ph,z}(\rho) \equiv \frac{1}{\rho_{ph}}\int \frac{d^3q}{(2\pi)^3} q\cos\theta_q n_{\vec q}\,(\vec r).
\label{vph}
\eeq
To first order in $T'$, only the $\cal D$ term in the distribution function survives the angular average in the numerator, so that
\beq
v_{ph,z}(\rho) =
 s \frac{T'}{T_0} \frac{ \int d^3q \cos^2\theta_q   q^2{\cal D} \,\partial n_q^0/\partial q }
{\int d^3q q^2 \,\partial n_q^0/\partial q},
\eeq
independent of $z$.
Since $\cal D$ is independent of $q$, the integrals over $q$ in numerator and denominator cancel, and 
\beq
v_{ph,z}(\rho) =  -3 s \frac{T'}{T_0} \langle {\cal D}\cos^2\theta_q \rangle,
\label{vphz}
\eeq
where the angular brackets denote the average over angles of $\vec q\,$.

    The flow velocity averaged over the cross section of the pipe (denoted by an overline) is simply
\beq
   \overline v_{ph,z}  = \frac{1}{\pi R^2} \int_0^R 2\pi \rho\, d\rho \, v_{ph,z}(\rho)    =  -3s  \frac{T'}{T_0}  {\cal D}_{eff},   \nonumber\\
\eeq   
where
\beq
   {\cal D}_{eff} = \frac{1}{\pi R^2} \int_0^R 2\pi \rho\, d\rho \, \langle {\cal D}\cos^2\theta_q\rangle.
   \label{deff}
\eeq
In evaluating $ {\cal D}_{eff}$, the integral over  $\theta_q$ decouples from those over $\rho$ and $\phi_q$, and the latter are easily performed if one integrates over $\rho$ before integrating over $\phi_q$.  One finds
\beq
 {\cal D} _{eff} =\frac{2R}{3}=\frac12 \overline{\langle   {\cal D} \rangle},
\label{dbar}
\eeq
which expresses the fact that the length important for thermal conduction is {\it one half\,} of $\overline{\langle   {\cal D} \rangle}$, the average distance to the wall of the pipe, averaged over the cross section of the pipe.   Since  $\overline v_{ph,z}= -3(K_{ph}/TC_{ph})T'$, where $C_{ph}$ is the phonon heat capacity, we find the phonon thermal conductivity,
\beq
  K_{ph}= s C_{ph}{\cal D}_{eff} = \frac23sC_{ph}R,
\eeq
which is the Casimir result, Ref.\ \cite{casimir}.    

\begin{figure}[h]
\includegraphics[width=8.5cm]{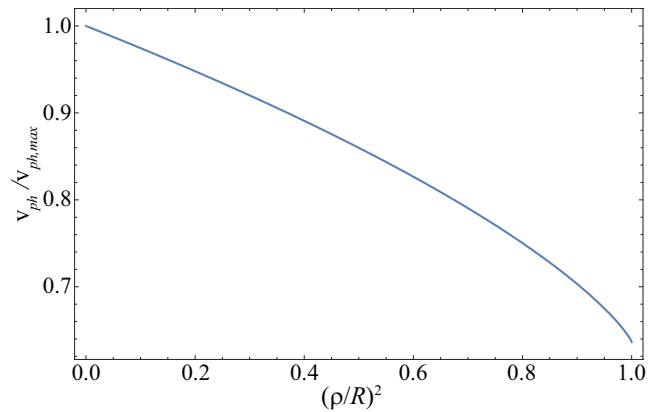}
\caption{Normalized velocity distribution across the cylinder as a function of $(\rho/R)^2$ in the Casimir limit.}
\label{fig:casimir}
\end{figure}

  Locally,
 \beq
\int \frac{d\Omega_q}{4\pi} {\cal D}\cos^2\theta_q =  \frac{\pi R}{4}I(\rho^2/R^2)
\eeq
where the elliptic integral,
\beq
I(t^2)=\frac{2}{\pi}\int_0^{\pi/2}d\phi\sqrt{1-t^2\sin^2\phi},
\eeq
must be done numerically.  As we see in Fig.~\ref{fig:casimir} the velocity profile
\beq
 v_{ph,z}(\rho) = -\frac{3\pi s  RT'}{4T_0}I(\rho^2/R^2),
\eeq
is independent of $z$ and nearly quadratic almost to the edge of the pipe, where it falls more rapidly, but unlike when viscosity dominates, it does not fall to zero at the pipe wall.

\section{Exact solution of the time dependent diffusion equation with advection }
\label{sec:AppendixB}

  Here we construct the general solutions of the time-dependent diffusion equation (\ref{diffeq}) by first transforming the equation
into self-adjoint form by writing $n_3(z,t) = e^{z/2h}\hat n(z,t)$.   As a result, $\hat n(z,t)$ obeys 
\beq
   \left(\frac{\partial }{\partial t} +  \frac{v_{ph}^2}{4D} - D \frac{\partial^2}{\partial z^2}\right)\hat n(z,t) =0,
   \label{diffeqhat}
\eeq
in the interval $0 \le z \le L$ (we simply write a generic diffusion constant, $D$, to simplify the notation).
The boundary condition of vanishing current at the two ends of the pipe then becomes 
\beq  
   \frac{\partial \hat n}{\partial z} =  \frac{\hat n}{2h} \quad\quad (z=0,L),
\eeq
where $h = D/v_{ph}$.   To realize the boundary conditions we expand $\hat n$ in a complete set of normalized solutions of Eq.~(\ref{diffeqhat})
that are periodic in the interval 0 to $2L$, 
\beq
   \phi_\nu(z) = \alpha_\nu \left(\cos kz + \frac{1}{2hk}\sin kz\right),
\eeq
with $k=\pi \nu/L$ ($\nu\ge1$), and 
\beq
\alpha_\nu =\left[(L/2)\left(1+1/(2hk_\nu)^2\right)\right]^{-1/2},
\eeq 
together with the stationary solution
\beq
\phi_0(z) = \frac{e^{z/2h}}{\left[h\left(e^{L/h}-1\right)\right]^{1/2}}.  
\eeq
The modes $\phi_\nu(z)$ form an orthonormal set obeying
\beq
  \int_0^L \phi_\nu(z)\phi_{\nu'}(z) dz = \delta_{\nu,\nu'}
\eeq
as well as the completeness relation in the interval $0 < z,z' < L$,
\beq
  \sum_{\nu=0}^\infty \phi_\nu(z)  \phi_\nu(z')  = \delta(z-z').
\eeq

The time dependence of the modes $\phi_\nu(z,t)$ is $e^{-t/\tau_\nu}$,
with 
\beq
\frac{1}{\tau_\nu}  =  \left( \frac{1}{4h^2} + k_\nu^2 \right)D, \quad \nu\ge 1,
\eeq 
and $1/\tau_0 = 0$.  Then
\beq
  \hat n(z,t) = \sum_{\nu=0}^\infty b_\nu \phi_\nu(z)e^{-t/\tau_\nu}
\eeq
with
\beq
  b_\nu = \int_0^L \hat n(z,0 )\phi_\nu(z) dz.
\eeq   

We can thus write the solution $\hat n(z,t)$ in terms of the initial density distribution as
\beq
   \hat n(z,t) = \int_0^L {\cal G}(z,z',t) \hat n(z',0)dz',
\eeq
where 
\beq
 {\cal G}(z,z',t) = \sum_{\nu=0}^\infty \phi_\nu(z)  \phi_\nu(z') e^{-t/\tau_\nu}\theta(t)
\eeq
is the Green's function for the diffusion equation in the form (\ref{diffeqhat}), with $\theta$ the Helmholtz unit step function.
At $t=0$, the completeness relation implies  ${\cal G}(z,z',t) = \delta(z-z')$, and thus
\beq
   \left(\frac{\partial }{\partial t} +  \frac{v_{ph}^2}{4D} - D \frac{\partial^2}{\partial z^2}\right){\cal G}(z,z',t) = \delta(z-z')\delta(t),
   \label{diffeqhat1}
\eeq
in the interval $0 \le z \le L$.

We now convert back to the original form of the diffusion equation, (\ref{diffeq}), and have
\beq
   n_3(z,t) = \int_0^L  G(z,z',t) \,n_3(z',0)dz',
   \label{dens}
\eeq  
where 
\beq
   G(z,z',t)& =& e^{(z+z')/2h}{\cal G}(z,z',t) 
   \label{greens}
\eeq 
is the Green's function for the original diffusion equation (\ref{diffeq}), i.e.,
\beq
   \left(\frac{\partial}{\partial t} +  v_{ph}\frac{\partial}{\partial z} - D \frac{\partial^2 }{\partial z^2}\right)G(z,z',t)  \nonumber\\
   =  e^{(z+z')/2h}\delta(z-z')\delta(t).   \nonumber\\
    \label{diffeqGG}
\eeq


\begin{thebibliography}{20}

\bibitem{hix} G. Baym, D. H. Beck, and C. J. Pethick, Phys. Rev. B {\bf 88}, 014512 (2013), arXiv:1212.2946.

\bibitem{BBPII} G. Baym, D. H. Beck, and C. J. Pethick, J. Low Temp. Phys., {\bf 178}, 200 (2015); arXiv:1408.1619.

\bibitem{BBP}   G. Baym, J. Bardeen and D. Pines,  Phys.  Rev. {\bf 156},
207 (1967); G. Baym and C. J. Pethick, {\it Landau Fermi liquid theory:  concepts and
applications,} (J.  Wiley and Sons,  New York, 1991).

\bibitem{snsExpt} R. Golub and S. K. Lamoreaux, Phys. Rep. {\bf 237}, 1 (1994).

\bibitem{hayden} M. Hayden, S. K. Lamoreaux, and R. Golub, AIP Conf. Proc. {\bf 850}, 147 (2006).

\bibitem{maris} H. J. Maris, Rev. Mod. Phys. {\bf 49}, 341 (1977).

\bibitem{BE67}  G. Baym and C. Ebner,  Phys.  Rev. {\bf 164}, 235 (1967).

\bibitem{optConc} The measurement is optimized when the neutron capture rate is approximately the same as the neutron decay rate---this occurs for concentrations $x_3 \sim 3\times 10^{-10}$.

\bibitem{BM} D. Benin and H. J. Maris, Phys. Rev. B {\bf 18}, 3112 (1978).
 
\bibitem{greywall} D. Greywall, Phys. Rev. B {\bf 23}, 2152 (1981).

\bibitem{casimir}  H. B. G. Casimir, Physica {\bf 5}, 495 (1938).    A basic assumption here is that phonons leaving the walls of the tube are in thermal equilibrium at the local temperature of the wall.  If there is significant specular reflection at the wall, the result is modified.

\bibitem{lamoreaux}  S. K. Lamoreaux, G. Archibald, P. D. Barnes, W. T. Buttler, D. J. Clark, M. D. Cooper, M. Espy, G. L. Greene, R. Golub, M. E. Hayden, C. Lei, L. J. Marek, J.-C. Peng, and S. Penttila, Europhys. Lett. {\bf 58}, 718 (2002).

\bibitem{knudsen} For wall scattering alone, the diffusion constant would be simply $D= {\bar v_3}\lambda/3$, where $\lambda$ is the average mean free path, here simply $2R$, the pipe diameter..


\bibitem{flush}  D. H. Beck et al. (to be published).

\end{thebibliography}
\end{document}